\documentclass[conference]{IEEEtran}

\usepackage{url}
\usepackage{fixltx2e}
\usepackage{amsmath,amssymb}
\usepackage{algorithm}
\usepackage{algorithmic}
\usepackage{flushend}

\usepackage{times,amsmath,epsfig}
\usepackage{epstopdf}
\usepackage{graphicx}
\usepackage{paralist}
\usepackage{subfig}
\usepackage{bm}
\usepackage{tikz}
\usepackage{pgfplots}
\usepackage{xcolor}
\usepackage{booktabs}
\usepackage{verbatim}
\usepackage{flushend}
\usepackage{tabularx}
\usepackage{hhline}
\usepackage{xcolor}
\usepackage{colortbl}

\usepackage{pgfplots}
\usepackage{multirow}
\usepackage{float}
\usepackage{listings}

\definecolor{dkgreen}{rgb}{0,0,0}
\definecolor{gray}{rgb}{0,0,0}
\definecolor{mauve}{rgb}{0,0,0}

\newcommand\JSONnumbervaluestyle{\color{black}}
\newcommand\JSONstringvaluestyle{\color{dkgreen}}

\newif\ifcolonfoundonthisline

\makeatletter

\lstdefinestyle{json}
{
  showstringspaces    = false,
  keywords            = {false,true},
  alsoletter          = 0123456789.,
  morestring          = [s]{"}{"},
  stringstyle         = \ifcolonfoundonthisline\JSONstringvaluestyle\fi,
  MoreSelectCharTable =%
    \lst@DefSaveDef{`:}\colon@json{\processColon@json},
  basicstyle          = \ttfamily,
  keywordstyle        = \ttfamily\bfseries,
}

\newcommand\processColon@json{%
  \colon@json%
  \ifnum\lst@mode=\lst@Pmode%
    \global\colonfoundonthislinetrue%
  \fi
}

\lst@AddToHook{Output}{%
  \ifcolonfoundonthisline%
    \ifnum\lst@mode=\lst@Pmode%
      \def\lst@thestyle{\JSONnumbervaluestyle}%
    \fi
  \fi
  \lsthk@DetectKeywords%
}

\lst@AddToHook{EOL}%
  {\global\colonfoundonthislinefalse}

\makeatother

\begin{document}
\title{Evaluation of the Performance of Adaptive HTTP Streaming Systems}
\date{}

\author{

\IEEEauthorblockN{Anatoliy Zabrovskiy, Evgeny Petrov,\\ Evgeny Kuzmin}
\IEEEauthorblockA{Petrozavodsk State University \\ Petrozavodsk, Russia \\ {\{z\_anatoliy, johnp, kuzmin}\}@petrsu.ru\\}

\and
\IEEEauthorblockN{Christian Timmerer}
\IEEEauthorblockA{Alpen-Adria-Universit{\"a}t Klagenfurt \\ Klagenfurt, Austria \\ christian.timmerer@itec.aau.at}

}
\maketitle

\begin{abstract}
Adaptive video streaming over HTTP is becoming omnipresent in our daily life. In the past, dozens of research papers have proposed novel approaches to address different aspects of adaptive streaming and a decent amount of player implementations (commercial and open source) are available. However, state of the art evaluations are sometimes superficial as many proposals only investigate a certain aspect of the problem or focus on a specific platform -- player implementations used in actual services are rarely considered. HTML5 is now available on many platforms and foster the deployment of adaptive media streaming applications.
We propose a common evaluation framework for adaptive HTML5 players and demonstrate its applicability by evaluating eight different players which are actually deployed in real-world services.


\end{abstract}

\section{Introduction}
\label{introduction}

Universal media access~\cite{SmithUMA} as proposed in the late 90s, early 2000 is now reality. It is very easy, in real-time to generate, distribute, share, and consume any media content, anywhere, anytime, and with/on any device. These kind of real-time entertainment services -- specifically, streaming audio and video -- are typically deployed over the open, unmanaged Internet and account now for more than 70\% of the evening traffic in North American fixed access networks. It is assumed that this number will reach 80 percent by the end of 2020~\cite{sandvine:2016}.

Using adaptive streaming techniques over HTTP is nowadays state of the art and massively deployed on the Internet adopting the over-the-top (OTT) paradigm, i.e., these services are deployed on top of existing infrastructures. For example, Netflix and YouTube alone account for more than 50\% of the traffic at peak periods~\cite{sandvine:2016}. 
Although Internet capacity is constantly increasing for both fixed and mobile networks, the adoption of new streaming services will continue as well as new applications and services will emerge.
Major formats in this domain are MPEG-DASH and Apple's HLS which both have the same underlying principles.

However, the current effort in MPEG referred to as Common Media Application Format (CMAF)~\cite{CMAF} aims at harmonizing at least segment formats and it is expected that soon DASH and HLS will support ISO base media file format (ISOBMFF) segments which are compatible with each other. In such a situation the most interest aspect -- at least from a research perspective -- is the rate adaption logic of players, because it is not defined in the standard and left open for competition.

In the past, many studies for such a rate adaptation logic were proposed and/or evaluated, e.g.,~\cite{Thang2014,TimmererMR2016}. However, most of them focus on the development of new adaptation algorithms and compare it only with a limited subset of existing approaches~\cite{Li2014,Huang2014,Yin2015,QoEStudy2017}. Evaluations of real-world deployments are very rare~\cite{Roverso2013}. Additionally, it is also difficult to come up with a comprehensive evaluation of existing approaches due to the lack of the appropriate tools.
Hence, the main contributions of this paper are as follows:
\begin{inparaenum}[(i)]
    \item we developed an adaptive video streaming evaluation framework for the automated testing of different players and, consequently rate adaptation logics;
    \item we identified eight well-known adaptive HTML5 players (commercial and open source) and integrated them in our framework;
    \item we conducted a series of experiments to prove the suitability and usefulness of our framework for the comparison of players and rate adaptation logics.
\end{inparaenum}
In general, in this paper we present a novel approach and framework for the automated evaluation of adaptive HTML5 players.

The rest of the paper is organized as follow. Related work is discussed in  Section~\ref{relatedwork}. Section~\ref{architecture} introduces the general architecture of our evaluation framework. An overview of the selected adaptive HTML5 players is given in Section~\ref{players}. The setup for the evaluation is described in Section~\ref{evalsetup}. Results are presented and discussed in Section~\ref{evalresults}. Section~\ref{conclusions} concludes the paper and highlights future work.

\section{Related Work}
\label{relatedwork}

Some time ago various rate adaptation algorithms have been presented and evaluated ~\cite{Thang2014,TimmererMR2016,Seufert2015,QoEStudy2017}. In general, most of the authors develop new algorithms and compare them only with a few existing rate adaptation algorithms or players~\cite{Li2014,Huang2014,Yin2015,QoEStudy2017}. Investigations of real adaptive HTML5 deployments have been found only in~\cite{Roverso2013}. 

The performance evaluation of adaptive streaming using network emulation have been addressed in prior studies ~\cite{Li2014,Huang2014,Maeki2015,Lederer2013}. Various approaches have been proposed and used for conducting experiment and adjusting bandwidth shaping trajectories. However, none of these studies use any automated and specially designed systems. 
In some cases bandwidth shaping is archived with Linux Traffic Control Program (tc)\footnote{\url{http://man7.org/linux/man-pages/man8/tc.8.html}} with the help of the special scripts or controlled by Linux Network Emulator (Netem)\footnote{\url{https://wiki.linuxfoundation.org/networking/netem
}}.

\section{System Architecture}
\label{architecture}

In this section we describe our system architecture enabling the automated evaluation of adaptive streaming systems within a controlled environment. Our proposed system comprises the following components:

\begin{itemize}
	\item Web server with standard HTTP hosting.
	\item Network emulation server.
	\item Selenium server.
	\item Web management interface. 
	\item Adaptive HTML5 players.
\end{itemize}

The system architecture is depicted in Figure~\ref{fig:systemarchitecture} and consists of the three servers running Ubuntu OS (version 16.04 LTS) and connected using Gigabit Ethernet switches.  
It defines a flexible system that allows adding new adaptive HTML5 players easily. There is an algorithm which describes the sequence of the steps of embedding a new player and its API. 

\begin{figure}[pt!]
\centering
\includegraphics[scale=0.5]{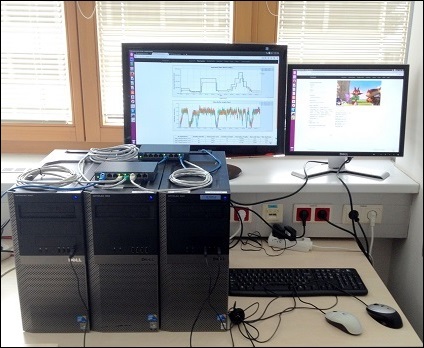}
\caption{Picture of the Server Infrastructure}
\label{fig:servers}
\end{figure}
\begin{figure*}[pt]
\centering
\includegraphics[scale=0.45]{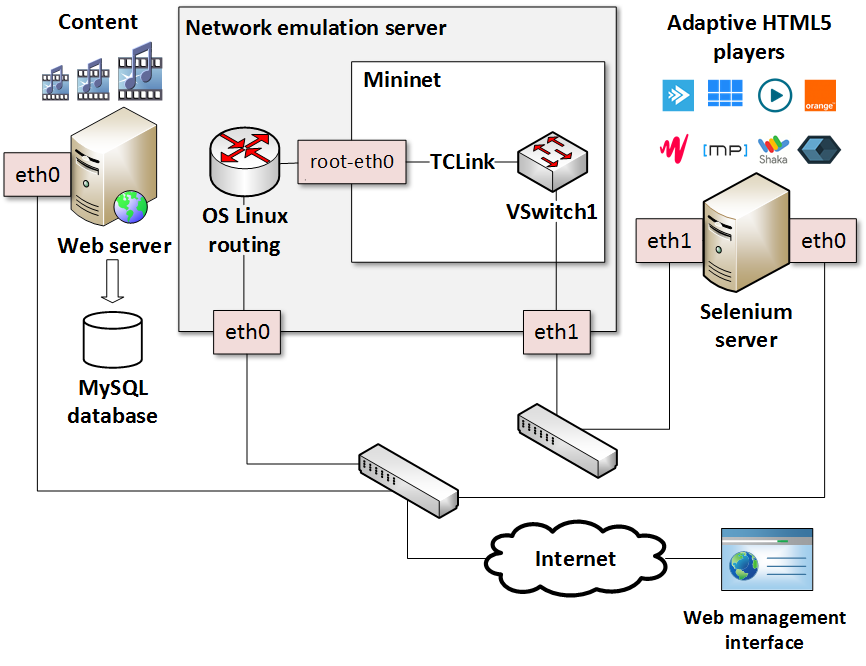}
\caption{System architecture}
\label{fig:systemarchitecture}
\end{figure*}

The \textit{Web server} hosts the video content for adaptive streaming over HTTP. 
For our experiments we adopted the MPEG-DASH format. Therefore, the video sequence will be provided in multiple configurations (e.g., bitrates, resolutions) which are referred to as representations. Each video sequence will be divided in segments of equal length measured in seconds of video content. Multiple versions of the same content, each version segmented in multiple smaller files. This enables the dynamic adaptation at segment boundaries according to the given context. 
It also hosts a MySQL database for collecting all the performance measurements and the \textit{Web management interfaces} for configuring and conducting the experiments. It is accessible from outside the controlled environment, everything else is within a controlled environment in order to avoid any cross-traffic that may influence the experiments. A picture of the real system is shown in Figure~\ref{fig:servers}.

The \textit{Selenium server\footnote{\url{http://www.seleniumhq.org/}}} is an open source software testing framework for Web applications which is used to automatically conduct our experiments with adaptive HTML5 players running within a Web browser. In our case we adopted the Google Chrome browser but it is also possible to use other browsers on various platforms (desktop, mobile, operating systems). The Selenium server is activated through the Web management interface to run the various experiments automatically according to a given configuration.

For the \textit{Network emulation server} we have adopted the Mininet\footnote{\url{http://mininet.org/}} emulator. Although this emulator is basically used for emulating  Software Defined Network (SDN) environments, it has been also used for streaming environments~\cite{MininetEmulation}. It provides a straightforward and extensible Python API for network creation and prototyping. We have utilized that functionality to create a virtual link with changeable network throughput characteristics. 
Our Network emulation server comprises two network interfaces (eth0, eth1). A python script is used to create a virtual network which consists of one switch connected to the real network using a TCLink\footnote{\url{http://mininet.org/api/classmininet_1_1link_1_1TCLink.html}}. The TCLink is a Mininet performance-modeling link. 
We setup our TCLink to change its characteristics at the specified moments of time. The schedule is stored within a file using JSON format. Finally, we made this schedule configurable through our Web management interface. 

The \textit{Web management interface} provides two functions,
\begin{inparaenum}[(i)]
  \item one for configuring and conducting the experiments and
  \item one which includes the player and provides real-time information about the currently conducted experiment.
\end{inparaenum}
Thus, the proposed framework in this paper provides means for comprehensive end-to-end evaluations of adaptive streaming services over HTTP including the possibility for subjective quality testing. The interface allows to define the following items and parameters:

\begin{itemize}
	\item configuration of network emulation profiles including the bandwidth trajectory, packet loss, and packet delay;
	\item specification of the number of runs of an experiment; and
	\item selection of the adaptive HTML5 player (or rate adaptation logics) and the utilized adaptive streaming protocol (MPEG-DASH or HLS).
\end{itemize}

The result page provides a list of conducted experiments and the analytics section contains various metrics of the conducted experiments. It is possible to generate graphs of the results by using Highcharts\footnote{\url{https://www.highcharts.com/}}
and export the raw values for further offline analysis. The following quality parameters and metrics are currently available:
download video bitrate; video buffer length; video startup time; stalls (or buffer underruns); number of quality switches; average video bitrate; instability and inefficiency~\cite{Jiang:2014}; simple QoE models specially designed for the adaptive streaming solutions~\cite{Maeki2015,Mok2011}.

Before starting the experiment we need to create a bandwidth trajectory profile. For each profile we can define duration of each stage, bandwidth, delay, and packet loss. As soon as we start an experiment within the Web management interface, the Google  Chrome  browser  (version  55.0.2883.87 64  bit) is automatically launched on the Selenium  server and the selected network profile including the link parameters is sent to Network emulation server. The actual requests for the video content towards the Web server goes through the Network emulation server.
When running an experiment it is possible to display the currently selected adaptive HTML5 player (including the video streaming) and real-time information about the currently conducted experiment. Details about the actual evaluation setup including all parameters and metrics are described in Section~\ref{evalsetup}.
All of the selected \textit{Adaptive HTML5 players} (cf. Table~\ref{table:players} in alphabetic order) have been available at no or relatively low costs for evaluation purposes.

\section{Overview of Adaptive HTML5\\Players}
\label{players}

In this section we provide a brief overview of the selected adaptive HTML5 players (in alphabetic order). All of them are implemented in Javascript and utilize the Media Source Extensions\footnote{\url{https://www.w3.org/TR/media-source/}} available on all modern browser platforms. In general, all players have their own API with methods, properties, and events which are used to obtain all metrics used in this paper. An overview of the player versions is shown in Table~\ref{table:players}. Please note that we are aware of the recently released THEOplayer\footnote{\url{http://www.theoplayer.com/}} but unfortunately it was not available at the time of writing this paper.

\subsection{Bitmovin Player}

The Bitmovin Player supports both MPEG-DASH and HLS and is freely available (up to 5,000 views). It provides a rich feature set including digital right management, advertisement, live streaming, and streaming of virtual reality and $360^\circ$ videos. It has a fully documented player API and various tutorials.

\subsection{dash.js}

The dash.js player is provided by the DASH Industry Forum (DASH-IF) and supports MPEG-DASH only. The DASH-IF is a non-profit industry association created to accelerate the proliferation of the MPEG-DASH standard. It provides interoperability guidelines, test vectors, conformance tools, and various software assets including the reference client dash.js.
 
The main idea of this project is to create and provide an open source JavaScript framework for using it in a real-world production environment. The player is free for commercial use and supports a wide range of options and features including a player API with full documentation. 

\subsection{Flowplayer}

The Flowplayer supports MPEG-DASH through a plugin architecture which is based on dash.js from the DASH-IF. This plugin gives full access to the dash.js player API via the $engine.dash$ property. It is expected to deliver a similar performance as dash.js unless the adaptation logic has been modified somehow.

\subsection{HAS Player}

The HAS Javascript player is also based on dash.js from the DASH-IF and available as an open source project. In addition to MPEG-DASH, it supports Microsoft Smooth Streaming and Apple HLS. Similar like the other players, a complete API documentation with the specification of public methods and events is available on the Web site of this project.

\subsection{JW Player}

The JW player is a commercial player which supports MPEG-DASH in Javascript and HTML5 since version 7. JW Player offers an entire video platform with different functionalities including a set of APIs for publishers and developers.

\subsection{Radiant Media Player}

Yet another commercial player supporting MPEG-DASH in HTML5 is the Radiant Media Player (MP). For client testing, the company offers a 14-day free trial version of the player with all features included. Interestingly, it is also based on dash.js and, thus, all methods and properties of the dash.js player can be accessed within Radiant MP. The API documentation includes an example of using the dash.js API with the Radiant MP.

\subsection{Shaka Player}

The Shaka player is offered by Google and is an open source JavaScript player supporting MPEG-DASH. The code of the Shaka Player needs to be compiled in order to be deployed on a Web site. The player API documentation includes methods and events for retrieving quality parameters and examples of how to use them.

\subsection{VideoJS Player}

Finally, the VideoJS player is a free and open source HTML5 video player which is also based on dash.js. Although the existing API does not provide methods and events for all required metrics, we have found a way to access these quality metrics through the instance of the dash.js which VideoJS player uses to support MPEG-DASH video playback.

\section{Evaluation Setup}
\label{evalsetup}

\begin{table}[pt!]
\caption{Overview of the adaptive HTML5 players.}\vspace{4pt}
\centering
\begin{tabularx}{0.48\textwidth}{|l|c|X|}
        \hline
        \textbf{Media player} & \textbf{Version} & \textbf{Web site (last access: \newline May 27, 2017)}\\
        \hline
        Bitmovin Player & 7.0 & \url{https://bitmovin.com}\\
        \hline
        dash.js & 2.4.0 & \url{http://dashif.org}\\
        \hline
		Flow Player & 6.0.5 & \url{https://flowplayer.org}\\
		\hline
		HAS Player & 1.7 & \url{https://github.com/Orange-OpenSource/hasplayer.js}\\
		\hline
		JW Player & 7.6.1 & \url{https://www.jwplayer.com}\\
		\hline
		Radiant MP & 3.10.8 & \url{https://www.radiantmediaplayer.com}\\
		\hline
		Shaka Player & 2.0.3 & \url{https://github.com/google/shaka-player}\\
		\hline
		VideoJS Player & 5.9.2 & \url{http://videojs.com}\\
		\hline
        \end{tabularx}
\label{table:players}
\end{table}

In this section, we define the setup for evaluating and comparing the adaptive HTML5 players. We describe what content is used and how it has been encoded, details about the network configuration, and the metrics used for the comparison. 

The MPEG-DASH content and a MPD file for our experiments have been produced using Bitmovin Cloud Encoding Service\footnote{\url{https://bitmovin.com/cloud-encoding-service/}}  by encoding the Big Buck Bunny animation movie\footnote{\url{http://bbb3d.renderfarming.net/download.html}} which is also a part of the DASH dataset~\cite{Lederer:2012}.The original video file characteristics are presented in the Table \ref{tab4}.

Note that our focus is primarily on the streaming performance, not a visual quality and, thus, we believe that one test sequence is sufficient.
We have encoded and prepared two different profiles as it is used in industry deployments. The first comprises a \textit{FullHD} profile with five different representations: 426x238 pixels (400kbps), 640x360 (800), 854x480 (1200), 1280x720 (2400), and 1920x1080 (4800). For the second configuration we reverse-engineered the \textit{Amazon} Prime video service which offers 15 different representations: 400x224 (100), 400x224 (150), 512x288 (200), 512x288 (300), 512x288 (500), 640x360 (800), 704x396 (1200), 704x396 (1800), 720x404 (2400), 720x404 (2500), 960x540 (2995), 1280x720 (3000), 1280x720 (4500), 1920x1080 (8000), and 1920x1080 (15000). In both cases we have adopted a segment length of four seconds as it provides a good trade-off regarding streaming performance and coding efficiency~\cite{Lederer:2012} which is also used in commercial deployments like Netflix.
The bitrate of audio stream for both sets was defined as 128 Kbps.

\begin{table}[pt!]
\centering
\caption{Original video file characteristics}\label{tab4}\vspace{4pt}
\begin{tabular}{|p{1.3in}|p{1.0in}|} \hline 
\textbf{\centerline{Video characteristic}} & \textbf{\centerline{Value}} \\
        \hline
        File size & 618 Mb\\
		\hline
		Frame rate & NTSC 30 fps\\
		\hline
		Resolution & 3840x2160 (4K)\\
		\hline
		Duration & 634 sec.\\
		\hline
		Average bitrate & 7498 Kbps\\
		\hline
\end{tabular}
\end{table}

\begin{figure}[ht!]
	\centering
	\begin{tikzpicture}
	\begin{axis}[
		title={Bandwidth Trajectory Scheme},
		xlabel={Time [Seconds]},
		ylabel={Bandwidth [kbps]},
		xmin=0, xmax=700,
    	ymin=0, ymax=4000,
   		xtick={0,120,240,360,480, 600, 720},
    	ytick={0,300,500,700,1000,1500,2000,2500,3500},
    	ymajorgrids=true,
    	grid style=dashed
	]
	\addplot[mark=square*, black, thick] coordinates {
		(0,750)
		
		(65,750)
		(65,350)
		
		(155,350)
		(155,2500)

		(275,2500)
		(275,500)
		
		(365,500)
		(365,700)
		
		(395,700)
		(395,1500)
		
		(425,1500)
		(425,2500)

		(455,2500)
		(455,3500)
		
		(485,3500)
		(485,2000)
		
		(515,2000)
		(515,1000)

		(545,1000)
		(545,500)
		
		(630,500)
	};
	\draw[gray, dashed, thick] (axis cs:630,\pgfkeysvalueof{/pgfplots/ymin}) -- (axis cs:630,\pgfkeysvalueof{/pgfplots/ymax});
	\end{axis}
	\end{tikzpicture}
	\caption{Bandwidth Trajectory Scheme used in the Experiments}
	\label{fig4}
\end{figure}
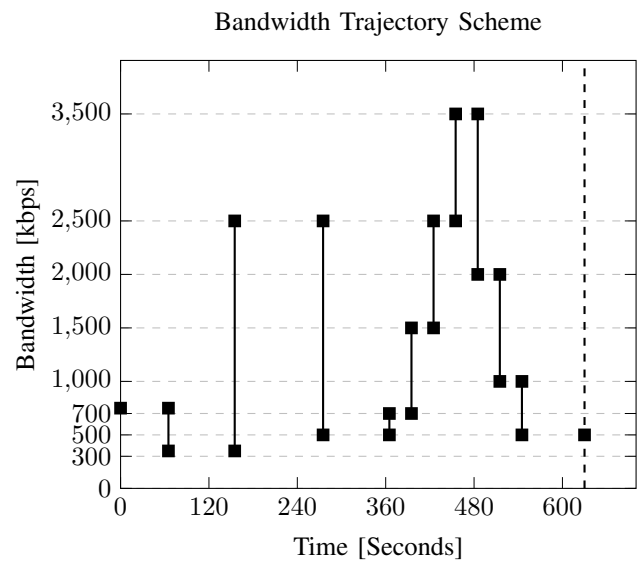

The \textit{network configuration} comprises a bandwidth trajectory adopted from~\cite{Muller:2012} providing both step-wise and abrupt adjustments in the available bandwidth to properly test all adaptive HTML5 players and its adaptation behavior under different conditions. The predefined bandwidth trajectory scheme is shown in Figure~\ref{fig4} and the bandwidth adjusts using the following sequence: 750 kbps (65 seconds), 350 kbps (90), 2500 kbps (120), 500 kbps (90), 700 kbps (30), 1500 kbps (30), 2500 kbps (30), 3500 kbps (30), 2000 kbps (30), 1000kbps (30) and 500 kbps (85). The network delay parameter was set to 70 milliseconds which corresponds to what can be observed within long-distance fixed line connections or reasonable mobile networks and, thus, is representative for a broad range of application scenarios. 

\begin{figure*}[pt]
\centering
\includegraphics[scale=0.63]{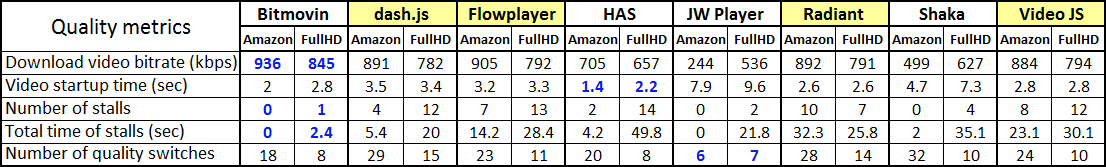}
\caption{Average results for Amazon and FullHD. Blue color indicates the best results and yellow indicates they might have the same adaptation logic}
\label{fig:averageresultstable}
\end{figure*}

\section{Evaluation Results}
\label{evalresults}

In this section, we present some results of our evaluation and discuss the key aspects. Each experiment was conducted five times and the average is presented here. We noted that the variance is quite low and, thus, we believe that five runs per experiment is sufficient. In total we conducted 80 experiments, each with a duration of 630 sec resulting in a total duration of 14 hours. In the figures of this section we present the results for both content configurations. The red bars refer to the $Amazon$ content configuration and the blue ones refer to the $FullHD$.

\begin{figure}[pt!]
\centering
\includegraphics[scale=0.7]{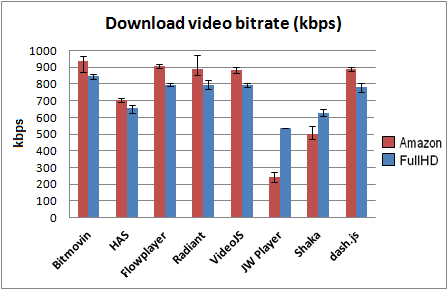}
\caption{Download video bitrate}
\label{fig:DownloadVideoBitrate_AmazonAndFullHD}
\end{figure}

Figure~\ref{fig:DownloadVideoBitrate_AmazonAndFullHD} shows that the Bitmovin player has the highest download video bitrate for both content configurations. The results of Flowplayer, Radiant MP, and VideoJS are very similar to dash.js as those players are based on dash.js and most likely adopt the same rate adaptation logic. 

\begin{figure}[pt!]
\centering
\includegraphics[scale=0.7]{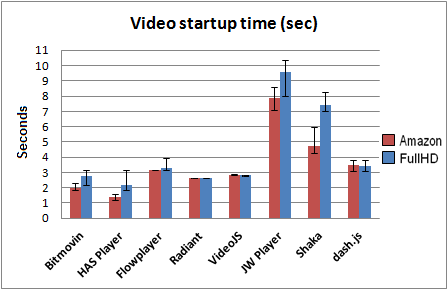}
\caption{Video startup time}
\label{fig:VideoStartupTime_AmazonAndFullHD}
\end{figure}

\begin{figure}[pt!]
\centering
\includegraphics[scale=0.7]{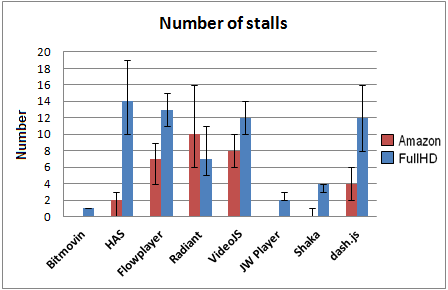}
\caption{Number of stalls}
\label{fig:NumberOfStalls_AmazonAndFullHD}
\end{figure}

The video startup time is shown in Figure~\ref{fig:VideoStartupTime_AmazonAndFullHD}. The results show that the HAS Player has the lowest video startup time and JW Player has the highest. Video startup time of Flowplayer, Radiant MP, and VideoJS is very similar to dash.js with minor differences. 
Interestingly, the number of stalls for the $Amazon$ profile is much lower than for the $FullHD$ profile as shown in Figure~\ref{fig:NumberOfStalls_AmazonAndFullHD} due to the fact that the former has a much higher number of content representations providing a better match to the bandwidth trajectory. The bandwidth trajectory has a reduction of the available bandwidth at the very beginning which does not match the lowest bitrate representation of the $FullHD$ profile and, thus, results in many stalls, at least for some players.

\begin{figure}[pt!]
\centering
\includegraphics[scale=0.7]{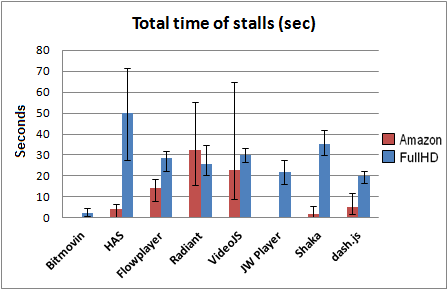}
\caption{Total time of stalls}
\label{fig:TotalTimeOfStalls_AmazonAndFullHD}
\end{figure}

\begin{figure}[pt!]
\centering
\includegraphics[scale=0.7]{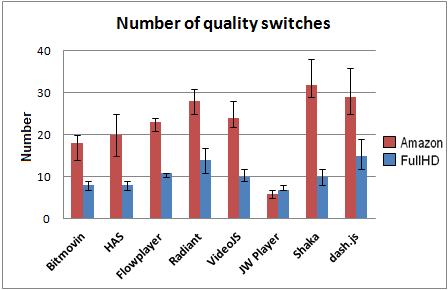}
\caption{Number of quality switches}
\label{fig:NumberOfQualitySwitсhes_AmazonAndFullHD}
\end{figure}

\begin{table}[pt!]
\centering
\caption{Instability and Inefficiency. Blue indicates best and red indicates worst}
\label{table:Instability_and_Inefficiency}
\includegraphics[scale=0.7]{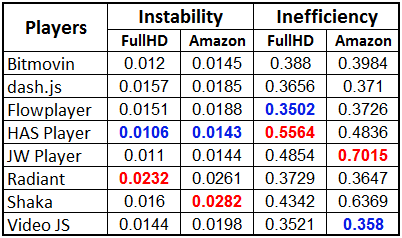}
\end{table}

The total time of stalls exposes additional findings about the adaptation behaviour of the different players. As shown in Figure~\ref{fig:TotalTimeOfStalls_AmazonAndFullHD} the Bitmovin player has the lowest total time of stalls for both profiles. Looking at the result for the $FullHD$ profile with five different content representations, the total time of stalls is \textgreater20 sec for all other players which may enormously affect the user perception. The results for the $Amazon$ profile are better due to the availability of more content representations. Nevertheless, the total time of stalls is still \textgreater20 sec for Radiant MP and VideoJS.

The number of quality switches is an important factor for the smoothness of the video streaming behaviour. As we can see in Figure~\ref{fig:NumberOfQualitySwitсhes_AmazonAndFullHD} it is twice as much for the $Amazon$ profile as more content representations are available. JW Player has a very low number of quality switches for the $Amazon$ profile but we also noticed that this player is inefficient with respect to bandwidth utilization (cf. download video bitrate).

The instability and inefficiency metrics for the $FullHD$ and the $Amazon$ profiles are shown in Table~\ref{table:Instability_and_Inefficiency}. Lower values of the instability metric reflect smoother video quality adaptation to the changing network characteristics. Lower values of the inefficiency metric indicate that the player rate adaptation algorithm more efficiently utilize the available network throughput in order to deliver the media content to the application. All tested players have fairly low values of instability indicating a smooth adaptation behaviour with a low number of quality switches. Only Radiant player has a bit higher instability and the Shaka Player has a higher instability but only for the $Amazon$ profile. In general, the number of quality switches using the $Amazon$ profile is twice as much as with the $FullHD$ profile which can be explained by the fact that the $Amazon$ profile has much more representations than the $FullHD$ profile. Inefficiency is comparable for the $FullHD$ profile except for the HAS Player which has a slightly higher inefficiency than the rest. Interestingly, JW Player and Shaka Player has a much higher inefficiency for the $Amazon$ profile -- both have the lowest download video bitrate -- and we again notice that Flowplayer, Radiant MP, and VideoJS provides a similar result as dash.js.

In general, we observe quite a different behaviour of all adaptive HTML5 players leading to different performance results. A summary of the results is provided in Figure~\ref{fig:averageresultstable}. All players adopt -- more or less -- a conservative approach according to the achieved download video bitrate. It seems that the Bitmovin player shows superior performance but also has the highest video buffer level (up to 40 sec) compared to all others (approx. 12-20 sec). Various user studies suggest that stalls should be avoided at all as they decrease the Quality of Experience (QoE) significantly. Looking at the results for the $Amazon$ profile, only the Bitmovin player, the JW Player, and the Shaka Player (with some outlier) achieve this goal. 

The evidence from these results suggests the following:
\begin{inparaenum}[(i)]
    \item in networks with bandwidth fluctuations, the playback quality significantly depends on the selected adaptive HTML5 player;
    \item it is reasonable to use suitable rate adaptation algorithms for different groups of users depending on the state of their network connections (e.g., dynamic switching of rate adaptation algorithms can be applied);
    \item the number of stalls depends on how many representations are available. The more bitrates exist, the lower the number of stalls.
\end{inparaenum}

\section{Conclusions and Future Work}
\label{conclusions}

In this paper we have evaluated eight adaptive HTML5 players which are actually and -- some of them -- massively deployed in the real-world services. 
In order to conduct the performance evaluations we developed an adaptive video streaming evaluation framework. With this framework it is possible to conduct a high number of experiments in a relatively short amount of time providing reliable results. The results of the experiments clearly show that the players have a different behavior depending on the status of the network characteristics and available content representations.
Future work will include adding new players (as they emerge on the market), investigating how different adaptive HTML5 players compete for the available bandwidth in a shared network and optimizing different adaptation algorithms depending on the network characteristics/conditions and the client devices.


\bibliographystyle{abbrv}
\vspace*{0.5mm}
\scriptsize
\bibliography{main.bbl}

\end{document}